\documentclass[manuscript,screen, authorversion]{acmart}

\copyrightyear{2024}
\acmYear{2024}
\setcopyright{rightsretained}
\acmConference[CUI '24]{ACM Conversational User Interfaces 2024}{July 8--10, 2024}{Luxembourg, Luxembourg}
\acmBooktitle{ACM Conversational User Interfaces 2024 (CUI '24), July 8--10, 2024, Luxembourg, Luxembourg}\acmDOI{10.1145/3640794.3665888}
\acmISBN{979-8-4007-0511-3/24/07}

\begin{document}

\title{Speculating About Multi-user Conversational Interfaces and LLMs: What If Chatting Wasn't So Lonely?}

\author{William Seymour}
\email{william.1.seymour@kcl.ac.uk}
\orcid{0000-0002-0256-6740}
\affiliation{%
  \institution{King's College London}
  \city{London}
  \country{UK}
}

\author{Emilee Rader}
\email{ejrader2@wisc.edu}
\orcid{0000-0003-2459-2744}
\affiliation{
    \institution{University of Wisconsin-Madison}
    \city{Madison}
    \state{Wisconsin}
    \country{USA}
}

\renewcommand{\shortauthors}{Seymour and Rader}

\begin{abstract}
The advent of LLMs means that CUIs are cool again, but what isn't so cool is that we're doomed to use them alone. The one user, one account, one device paradigm has dominated the design of CUIs and is not going away as new conversational technologies emerge. In this provocation we explore some of the technical, legal, and design difficulties that seem to make multi-user CUIs so difficult to implement.
Drawing inspiration from the ways that people manage messy group discussions, such as parliamentary and consensus-based paradigms, we show how LLM-based CUIs might be well suited to bridging the gap. With any luck, this might even result in everyone having to sit through fewer poorly run meetings and agonising group discussions---truly a laudable goal!
\end{abstract}



\maketitle

\section{Introduction}
With the meteoric rise and public uptake of LLMs such as ChatGPT, CUIs are suddenly back in the spotlight. Chatbots are cool again, and even your neighbours understand your awkward small talk about what you do for a living!

But something stands out about all of the major use-cases for the CUIs that we see around us. They're all quite... lonely. When we pour out our problems to a customer service chatbot, tell Google Assistant where we parked the car, or ask Gemini for the terminal velocity of a dog on a skateboard, it's just us. We're alone with the interface and nobody else is allowed to join in. We're so habituated to the one-user-one-device paradigm that we don't even notice it; if you try to imagine an LLM or similar interface that supports multiple users, you probably default to a scenario where two people use the CUI one at a time. Even voice assistants\footnote{RIP Google Conversational Actions, 2016-2023.}, \textit{devices that are supposed to be used in communal settings}, operate this way. The first person to speak starts the request, and anything that's said by others is jumbled in with it, like it's inconceivable that there might be other people using the conversational interface or just in the same room. Yes, there are stereotypes about computer science majors, \textit{but this is just ridiculous!}

The serious point here is that these technologies are being pitched as the future of how we work on just about everything, with video demos of people working (alone) with conversational AI in order to create documents, spreadsheets, visualisations and travel itineraries. But at the same time as we gain a collaborator in an LLM, we lose the ability to work with any humans we might want or need to be on board. This isn't how people work in the real world. ChatGPT might be able to help you draft an email, but it's not geared up to support a team working together. In collaborative tasks like creating documents and presentations, we can't help but chat; Google Docs, Overleaf, and Office Online all have the ability to chat in real time because they're central to how people work together. Indeed, if you want to make a point that's overly complex to write down, the default option is to have a meeting!

\section{So how did we get here?}
Multi-user interfaces in general are \textit{messy}. Keeping track of what different people are doing, resolving conflicts over the state of different users' interactions, and dealing with simultaneous inputs present difficult design and engineering challenges that disappear under the assumption that there will only be one user at a time (and by extension, that only one user will be logged in at a time). Some of these apply doubly so to voice; for example, while it's possible to reliably distinguish between speakers, systems often don't, and it is not possible to reliably separate out speech from people talking over each other. This arises in the literature where we see `competitions' over the use of Alexa~\cite{10.1145/3173574.3174214} due to what is often generously described as the ``equal access'' afforded by its single speaker assumption~\cite{beirl2019using}, or problems around collecting consent when it is not clear who is giving that consent~\cite{10.1145/3544548.3580967}.

Additionally, from a compliance perspective we see that most data protection regulations operate (broadly speaking) on the basis of protecting data about a single data subject from a single external data controller. When faced with the reality that more than one organisation might be involved, the GDPR pushes them towards being data \textit{processors} instead. Where joint controllership is unavoidable, they are treated like one big controller with joint liability. There is also no real provision for data generated by a group. This shapes the design of devices and services, leading to everything being built on a one-to-one model between companies and end users.

The same difficulties arise in group discussions in real-world settings: multiple participants mean they're messy! But people have developed ways of managing the mess---think of a parliament or assembly scenario in which over a hundred people with differing goals come together to debate and make decisions with a human mediator or `speaker'. People similarly manage to communicate in online chat rooms that have multiple people~\cite{mcculloch2020because}. \textbf{This causes us to ask the question: how might we take inspiration from these situations to generate better multi-user CUIs that can potentially scaffold interactions with many people at once?} To that end, in this paper we advance the discussion beyond the idea of multi-user CUIs as technologies that are simply aware of multiple interlocutors. Instead, we speculate about how they might support situations where a group is tasked with making a difficult or contentious decision.

\section{Exploring Alternative Paradigms}
Without the exploration of alternative paradigms, there's a real risk that our conversations with CUIs will continue to be \emph{shaped} by the technology, instead of \emph{supported} by it. What's happening in smart homes is a good example of this, with a general awareness that, contrary to how they are designed and marketed, `personal' devices are often anything but~\cite{10.1145/2858036.2858051}. A similar amount of wishful thinking takes place around the concept of family; Goulden uses the term \textit{platform family} to refer to the ``engineered simulacra of domesticity, formatted to run on the smart home operating system, serving simultaneously as a vehicle for domestic consumption, and a vehicle for consuming domestic life''~\cite{10.1080/1369118X.2019.1668454}. You might have come across something similar yourself if you've looked at `family' account options that only accommodate two adults living at the same address, exactly two children, and only distinguish between `adults' and `children'. In situations where we've become so habituated to a way of doing things that we lose the ability to think outside the box and start to unthinkingly accept the status quo, we have to look elsewhere for inspiration as to how we organise our speech and conversation! 

One challenge in group decision-making is that different people have access to different knowledge and information, and generally better group decisions are made when individuals share their information and preferences as part of the decision-making process~\cite{ErvinKeyton2019}. So there must be some way for people to discuss and reach an agreement on how to proceed. Differences in perspectives and disagreement are an important part of this, because without disagreement there would be no discussion; essentially, disagreement and positive interactions around discussing different perspectives leads to better decisions, because it gets people to talk and share ideas and information~\cite{ErvinKeyton2019}. However, this disagreement often makes group discussions hard to manage. Below we describe two different ways of organising group discussion and decision-making in an attempt to spark some ideas for creative ways to build truly multi-user conversational experiences around the new large language models that have recently stolen the spotlight.

\section{Parliamentary Systems}
A classic example of a system that organizes discussion on contentious issues is the `parliamentary' process of making decisions used in communities and organisations around the world. As one might expect, prominent examples can be found in parliaments where there is (in theory!) a focus on debate and a requirement that decisions receive the support of the majority. But this also includes similar systems found on smaller scales in committee meetings of various types, and we use ``\textit{Roberts Rules of Order}'' (``Robert's Rules'', originally designed to keep order during decision making in the US army) as a point of reference~\cite{robert2020robert}. Broadly speaking:

\begin{enumerate}
    \item Somebody proposes a \textbf{motion} describing a desired rule or action
    \item The group debates that motion
    \item Parts of the motion can be \textbf{added}, \textbf{substituted}, or \textbf{removed}
    \item The group \textbf{votes} on the motion and its potential amendments, and requires a threshold of `yes' votes to pass
    \item Motions can later be \textbf{amended} or \textbf{repealed}
\end{enumerate}

Here people come together with their own goals and try to convince others of the merits of their approach according to an established set of rules that can themselves be modified using those same rules. Having rules about who can speak at different points minimises crosstalk and other disruptions that can steer the conversation away from the issue being decided. Robert's Rules values process---that is, following the rules. The rules help to organize discussion and decision-making so that progress can be made, even when people disagree. We are used to the moderator (or ``speaker'') in these situations being a human, but given that the core of the role is interpreting what people say and responding according to the rules there is no reason in principle that this could not be performed by a CUI. 

\subsection*{Scenario 1: The Departmental Meeting}
This scenario takes place in the monthly departmental meeting of a typical university informatics department. The head of department has been informed by senior management that the university needs to reduce the amount of office space it occupies. As a result, they will be exploring ways to collect data about desk utilisation to aid decision-making about which spaces to keep. Facial recognition cameras have been proposed as a way of accurately tracking how many people are using different parts of the building. Cognisant of how potential privacy and trust violations could cause problems for the daily operation of the department and the possibility that people might try and circumvent the cameras, the head of department has convened a meeting to discuss the implementation plan. The online meeting is of about 50 academics and is moderated by a CUI:

\begin{itemize}
    \item (CUI) This meeting is to discuss the new occupancy tracking plan. I've muted everyone, please press ``raise hand'' to speak.
    \item (CUI) Senior management has proposed a scheme whereby we will use facial recognition cameras to track occupancy, with offices used by less than 50\% of their occupants, averaged over a month, to be reassigned.
    \item (\textit{people raise their virtual hands and are given 2 minutes to voice their opinion when called upon by the CUI...})
    \item (CUI) To summarise, it seems like we have two schools of thought: 1) that the introduction of the cameras is a way for senior management to surreptitiously check how long we spend working; and 2) given that there are some underutilised spaces, that motion detectors would be a more privacy preserving way of achieving the same goal.
    \item (CUI) Given this, I propose two corresponding motions: 1) that the department calls on senior management to categorically state that they are not using the scheme to track colleagues' working hours; and 2) that the department proposes the use of motion detectors instead of facial recognition cameras to senior management. Please raise your hand if you wish to discuss or amend the first proposal.
    \item (\textit{people raise their virtual hands and are given 2 minutes to voice their opinion when called upon by the CUI...})
    \item (CUI) Thank you, that concludes the discussion. Everyone in favour of the motion that ``the department refuses to engage in...'' please raise your hand
    \item (CUI) The majority was opposed, so the motion has not passed.
    \item [] {[...]}
    \item (CUI) That's the last item on the agenda for today. Please raise your hand if there's any other business, otherwise the next meeting is on ...
\end{itemize}

While we don't actually expect to see department meetings around the world suddenly be taken over by CUIs---that would probably get very annoying very quickly---this scenario illustrates how CUIs could play a role in mediating debates and decision making. This isn't as much of a leap as it might seem: we already see bots on platforms like discord that oversee more simplistic voting and decision making, and Copilot is already being used to generate meeting summaries. We've all struggled through meetings with inexperienced chairs---perhaps a system like this could give them nudges about when to curtail a discussion, improve the quality of minutes, or answer context-sensitive questions on procedure? 

\section{Consensus Building}
Another important way people organise group deliberation and decision-making is through consensus building. Consensus is what occurs when a decision is reached by ``a general or widespread agreement among the involved people''~\cite{Perez2018}. Consensus is usually reached through discussion and negotiation, and often the parties involved change their position and opinions as a result of hearing about others' perspectives, along the way to reaching a decision that is acceptable to all.

An advantage of consensus models over voting based processes like Robert's Rules is that because an agreement has been reached, it is assumed that everyone will accept the decision and go along with it. For this to be true, everyone involved in the decision must participate in the discussion, and their perspectives must be heard and addressed to the satisfaction of all. What kind of mediated experiences could we facilitate where a CUI encourages less vocal participants to engage, summarises the different positions, and even suggests areas of similarity and compromise?

\subsection*{Scenario 2: Who can see the photo?}
Ada, Max, Sam, and Chi went out to a club and took a photo together. Once Ada uploads it to social media, the platform begins the negotiation process whereby they must reach a consensus on how the photo will be shared before it becomes visible using a text-based CUI:

\begin{itemize}
    \item (Ada) I'm happy to share the photo with everyone.
    \item (Max) Can we just share it with our mutual friends? I don't want my parents to see it.
    \item (Sam) But I want to share it with my colleagues as well.
    \item (CUI) Maybe I could share the photo with \textbf{your mutual friends} and \textbf{Sam's colleagues}?
    \item (Chi) Actually, I'm not really comfortable with anyone that I don't know seeing the photo.
    \item (CUI) Okay, no problem. We can't satisfy everyone with one photo, but maybe I could share this version of the photo with \textbf{your mutual friends} and a cropped version without Chi with \textbf{your individual friends and Sam's colleagues}?
    \item (Ada, Chi, Max) That works!
\end{itemize}

In these situations, it currently tends to be whoever took (or `owns') the photo who unilaterally decides who can see it and what to do with it. Here, however, the group has decided that they will reach a collective agreement and the CUI assists them by presenting different options that satisfy most or all of their requests. An important part of the CUI's role in this situation is listening to the views of unhappy stakeholders and attempting to accommodate them. And again, the groundwork has already been laid by an abundance of contributions in the literature that attempt to solve this kind of problem algorithmically, given various information about the stakeholders and their preferences~\cite{10.1145/3146025, 10.1145/3360498, 10.1145/3208039}.

\section{What Values Do These Paradigms Embody?}
An important question to ask about the different ways that we interact with CUIs, whether now or (as here) in the future, is what values different interaction paradigms embody. The one-user-one-account paradigm, for example, limits the input to a single person. The contributions of other people involved in the discussion are at best, secondary and at worst, ignored. This authoritarian approach emphasises the power and control of one member of the team, group, or household who is usually the most technically engaged (and also usually a man). A CUI in this paradigm is essentially a filter, enabling some to participate while others can't. 

Parliamentary processes place high value on majority rule in order to ensure progress in the time allotted. This means it is possible to have a decision that is reached through majority support that still has a large minority that is unhappy with the decision. If that happens, the decision may be undermined when the minority who voted against it refuse to go along with it, or are vocally unhappy about it. A CUI could help a contentious meeting run more smoothly and result in a better decision because it doesn't have a position or a stake in the outcome. It may be viewed as more impartial as it works to ensure that all voices have a chance to be heard. This impartiality is not always successfully enacted in practice by the chair of a meeting following Robert's Rules; while chairs are supposed to run the meeting but not participate in it, personal positions and biases can affect how they run the meeting. A CUI in this interaction paradigm could essentially be a parliamentarian, making sure the meeting runs smoothly and forming the motions so that the attention of the meeting participants is on the issues. This supporting role means the CUI is a facilitator that does not have a `voice' or vote in the meeting, and its only influence on the outcome is in ensuring that the meeting participants are able to contribute their perspectives and vote.

Consensus based approaches place value on involvement and agreement from all parties. One problem with this is that reaching consensus can take a long time, and it requires active participation and compromise from all to be true consensus. There is also some flexibility around the extent to which all parties must be in agreement versus all parties having their views heard and considered. This can make discussions take longer and be more effortful while people unpack their gut feelings and form preferences discursively (although this is in itself something CUIs are well placed to help with!). For contexts like privacy, decisions made in the abstract may also not match what people want to happen in a specific situation, making it possible that certain decisions reached by consensus may not be the right decision or even a good one in the end. A CUI in this paradigm would be closer to an active meeting participant, attending to the concerns of all participants and suggesting potential ways forward that take their views into account. 

\section{Conclusion}
In 2000, Ackerman wrote about the social-technical gap: ``the divide between what we know we must support socially and what we can support technically''~\cite{Ackerman2000}. LLMs provide an opportunity to bridge this gap through support for group discussion and decision-making. These two scenarios illustrate that maybe we \textit{could} avoid the trap of the one-user-one-device model. What if the work of managing social situations didn't have to be pushed onto the user, and the system could provide support for this?

An opportunity exists for CUIs to get involved and make the process better by filling roles that can be difficult for human participants. Still, we must be careful to consider when such roles for CUIs might produce positive outcomes, and when they might not. For example, a CUI serving as a parliamentarian or consensus facilitator could sow chaos, intentionally or unintentionally, if the system were compromised or manipulated to produce negative outcomes---or even if it were just badly designed. Also, having a CUI fill a role that is currently held by a human is a step that should not be undertaken lightly. Delegating the roles described above to CUIs risks having one's voice overruled by a large language model. This would appear to substantiate fears that CUIs might ``systematically [interfere] with human professional relationships by way of their elevated positions within a social hierarchy''~\cite{10.1145/3357236.3395479}. As with the platform family~\cite{10.1080/1369118X.2019.1668454}, the deployment of these technologies in elevated positions, particularly where they are imposed unilaterally, could result in homogenisation and the gradual stamping out of non-conformity. Careful consideration of the values that underlie the different interaction paradigms (authoritarian, process-focused, consensus-focused) is an important first step towards ensuring that these systems are developed to support and facilitate messy group discussions, not eliminate them.

However, conversational interfaces don't have to do everything. They could provide some scaffolding for discussion without being rigid or telling people what to do, fitting into human communication processes in such a way that they can play supporting roles that make for better and more efficient decisions through encouraging participation, organising contributions, and summarising positions. The key thing is that we try to break out of the restrictive design patterns that we've become stuck in, so that we're not perpetually alone talking to ChatGPT.

\bibliographystyle{ACM-Reference-Format}
\bibliography{main}
\end{document}